\documentclass{appolb}
\usepackage{graphicx}
\usepackage{hyperref}
\usepackage{subfigure}
\usepackage{mathrsfs}
\usepackage{cancel}
\usepackage{bbm}

\begin{document}
\title{The quark propagator in the complex domain: A status report%
\thanks{Presented at Excited QCD 2017, Sintra, Portugal}%
}
\author{Andreas Windisch
\address{Department of Physics, Washington University in St. Louis, MO 63108, USA}
\\
}
\maketitle
\begin{abstract}
In these proceedings I review the status of an ongoing project that aims at solving the quark propagator Dyson-Schwinger equation in the complex domain. The novel aspect of the approach is that the non-analyticities arising throughout the iteration of the equation are to be taken into account in an appropriate way through dynamic contour deformation. Because of the complexity of the approach, these studies are undertaken in a heavily truncated scenario that serves as a toy model for the development of the numerical techniques that are required to treat the system in a mathematically sound way.
\end{abstract}
  
\section{Introduction}
The analytic properties of the quark propagator Dyson-Schwinger equation (DSE) were in the focus of many studies \cite{Stainsby:1992hy,Bender:1994bv,Burden:1997ja,Alkofer:2003jj,Fischer:2008sp,Eichmann:2009zx,Krassnigg:2009gd,Dorkin:2013rsa,El-Bennich:2016qmb,Windisch:2016iud}, to present a non-exhaustive list that spans more than two decades. While most of these studies were conducted in Euclidean space, there are also investigations in Minkowski space, e.g.~\cite{Sauli:2013vsa}, as well as other impressive studies of the analytic properties of various Green's functions using DSEs \cite{Fischer:2009jm,Strauss:2012dg}. The interest in this field is mainly driven by the necessity of extending the real quark propagator solutions to complex momenta as required for bound state equations (see e.g. \cite{Sanchis-Alepuz:2015tha}), but also because the positivity properties of the propagator can be used to establish a sufficient criterion to remove a certain degree of freedom from the space of asymptotic states (see e.g. \cite{Alkofer:2003jj}). Thus, knowing the solution of the quark propagator DSE in the complex domain is highly desirable. Unfortunately, the evaluation of the quark DSE for complex external momenta is tedious. The main complication is, that once the external momentum is allowed to be a complex number, the integration contour of the quark self-energy loop must be deformed away from the real axis, such that the loop momentum has to be treated as a complex quantity as well.

\section{Building a reliable numerical framework}
The rainbow truncated quark propagator DSE reads
\begin{eqnarray}
S^{-1}\left(p\right) & = & S_{0}^{-1}\left(p\right)+\int\frac{d^{4}q}{\left(2\pi\right)^{4}}\left[\mathcal{G}\left(\left(p-q\right)^{2}\right)\right.\nonumber \\
 &  & \times\left.\left(p-q\right)^{2}D_{free}^{\mu\nu}\left(p-q\right)\gamma^{\mu}S\left(q\right)\gamma^{\nu}\right],\label{eq:7}
\end{eqnarray}
where the inverse propagator is given by
\begin{eqnarray}
S^{-1}\left(p\right) & = & \delta^{\alpha\beta}\left(i\cancel{p}\ A\left(p^{2}\right)+B\left(p^{2}\right)\mathbbm{1}_D\right)\label{eq:3},
\end{eqnarray}
and the interaction term $\mathcal{G}$ is specified in Section \ref{inclUV}.
Solving the quark propagator DSE in the complex domain requires fast, reliable and robust numerical tools.
\begin{figure}
\centering
\includegraphics[width=8.5cm]{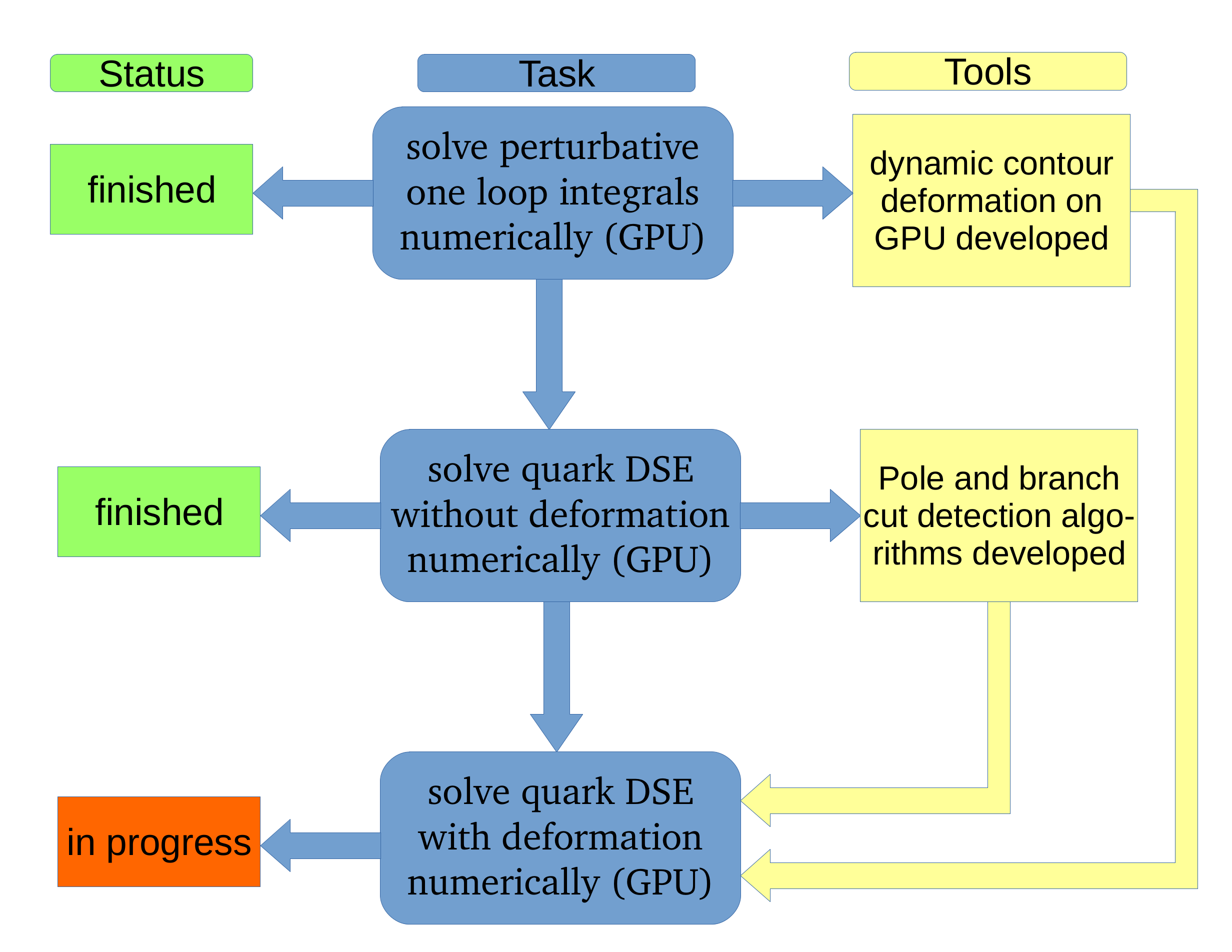}
\caption{Outline of the project, see main text.}
\label{Fig:1}
\end{figure}
Establishing such a framework will be achieved through the project  outlined in Figure \ref{Fig:1}. It is partitioned into three sections.
\begin{itemize}
\item\textbf{Perturbative one loop integrals in the complex domain}\\
(see references \cite{Windisch:2012zd,Windisch:2012sz,Windisch:2013dxa})\\
In these studies, the problem of solving Green's functions to one-loop order in the complex domain has been addressed. While these calculations were simple in the sense that there was no iterative procedure involved, they featured all the complications induced by the external momentum being complex. Allowing the external momentum to be complex leads to branch cuts in the complex plane of the radial integration variable of the loop integral, and the contours have to be deformed dynamically. This approach has been implemented on a Graphics Processing Unit (GPU), and the results have been verified by studying an example for which an exact solution is known. Since all quantities (and their analytic properties) in the loop integral are known, one can determine the analytic properties of the integrand prior to the integration, and choose appropriate contours depending on where the obstructive structures appear. Similar techniques to the ones used here have been recently applied successfully to a complicated integrand with two overlapping branch-cuts \cite{Weil:2017knt}. 

\item\textbf{The quark DSE for complex momenta without contour deformation}\\
(see reference \cite{Windisch:2016iud})\\
The second important step towards building a framework that can cope with the complications arising in the context of complex external momenta is to study iterative integral equations for complex momenta without deforming the contour. In the case of the quark propagator, there are truncations that give rise to analytic integrands for the quark propagator dressing functions (IR part of \cite{Maris:1999nt}, as well as \cite{Alkofer:2002bp}), and no contour deformation is required. I used this setting to develop tools for detecting poles in the solution numerically. 

\item\textbf{The quark DSE for complex momenta with contour deformation}\\
(in progress)\\
The last step in this program is to solve the full case, that is, adjust the integration contours as necessary. The main difference to the perturbative case discussed above is, that the analytic properties of the (self-consistently obtained) solution is not known a priori. However, prior to solving the equation iteratively one has to choose initial values for the complex dressing functions $A$ and $B$. Once the initial functions have been chosen, the situation after the first iteration step is similar to the case of the perturbative integration as discussed in \cite{Windisch:2013mg,Windisch:2013dxa}. Using the tools (with some extensions) for pole detection discussed in \cite{Windisch:2016iud}, the code that iterates the system in the complex plane can dynamically produce a contour deformation that is appropriate for the analytic properties of the integrand at any given iteration step. Since the contours for the first iteration step can be worked out by hand easily, one can also test the system by comparing the deformations produced by the code with the ones that have been worked out explicitly. Using this technique, in principle any integration kernel that does not introduce further unknown quantities (for example, dressing functions of the quark-gluon vertex without knowledge of their analytic properties) can be treated in a mathematically sound way. Apart from the Maris-Tandy model used here, one can then, for example, study the Qin-Chang model \cite{Qin:2011dd}, or even go beyond the rainbow approximation. 
\end{itemize}
Following these three steps, I hope that I will be able to provide a fast, reliable and robust numerical framework to solve the quark propagator DSE in the complex domain.

\section{\label{inclUV}Adding the UV term without contour deformation}
Before studying the contour deformations in detail, it is of course interesting to see what happens to the pole structure if the ultra-violet term of the Maris-Tandy model is taken into account, while the contour is maintained along the real axis. The Maris-Tandy model is given by
\begin{eqnarray}
\label{eq:MT_model}
Z_{1F}\ g^{2}\frac{\mathscr{G}}{k^{2}} & = & \frac{4\pi^{2}}{\omega^{6}}Dk^{2}e^{-\frac{k^{2}}{\omega^{2}}}+4\pi^{2}\frac{\frac{12}{33-2N_{f}}\frac{1}{k^{2}}\left(1-e^{-\frac{k^{2}}{4m_{t}^{2}}}\right) }{\frac{1}{2}\ln\left[e^{2}-1+\left(1+\frac{k^{2}}{\Lambda_{QCD}^{2}}\right)^{2}\right]},
\end{eqnarray}
and the parameters used here are $N_{f}=4$, $\Lambda_{QCD}^{Nf=4}=0.234$
GeV, $\omega=0.5$ GeV, $D=1.0$ $\mbox{GeV}^{2}$ and $m_{t}=0.5$
GeV.
\subfiglabelskip=0pt
\begin{figure*}
\centering
\subfigure[][]{
 \label{fig:2_a}
\includegraphics[width=0.45\hsize]{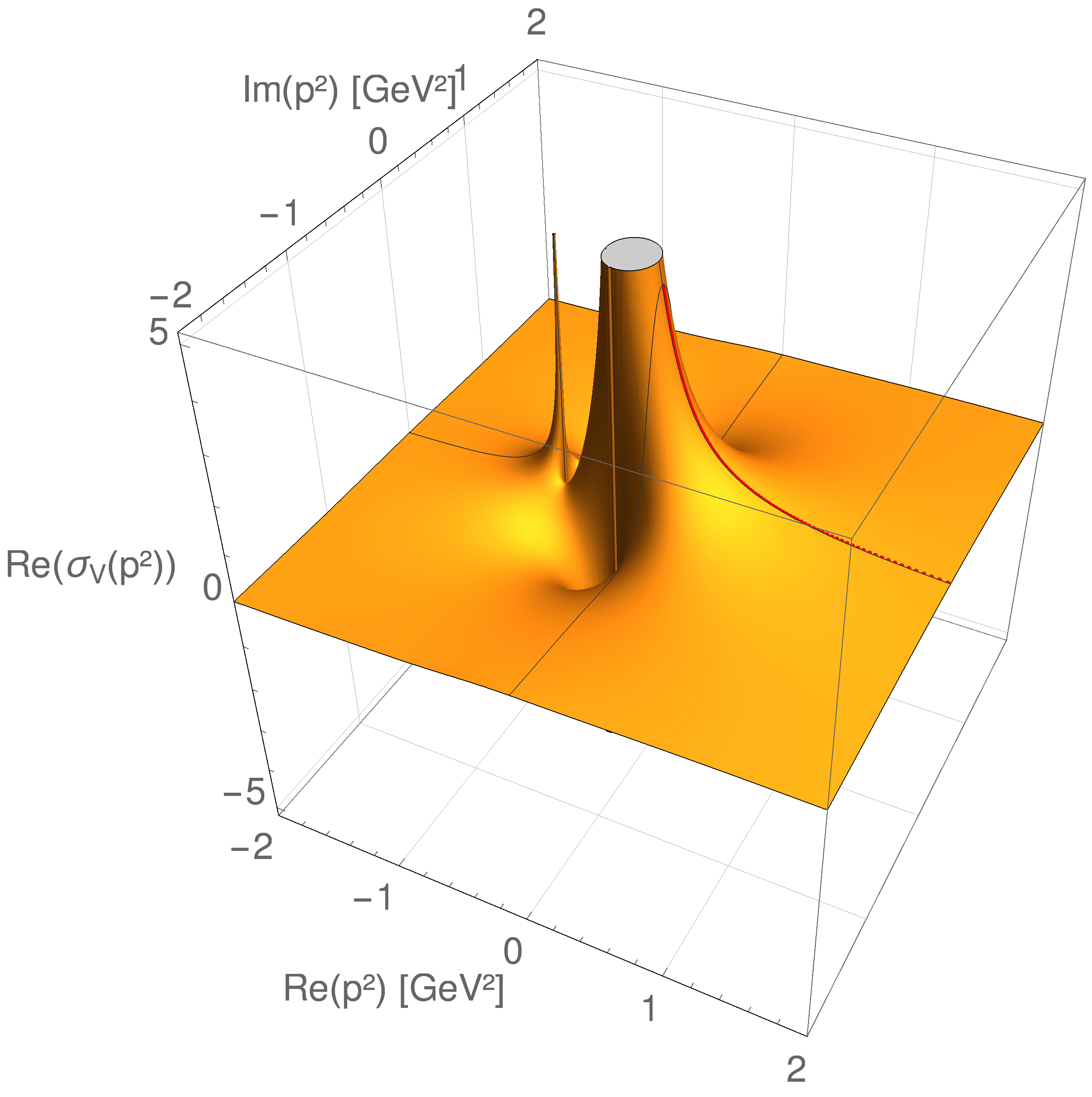}
}\hspace{8pt}
\subfigure[][]{
 \label{fig:2_b}
\includegraphics[width=0.45\hsize]{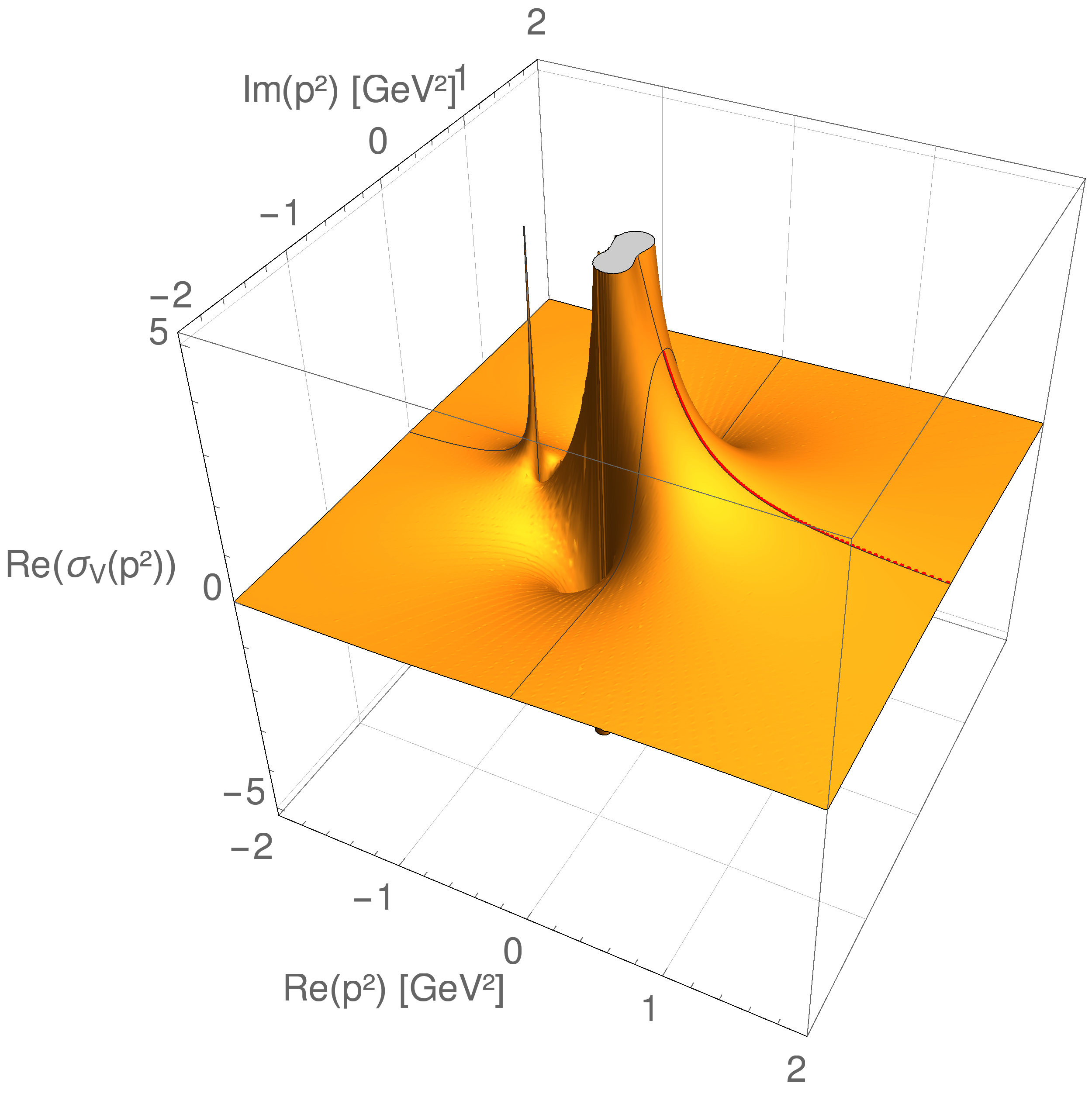}
}
\caption[]{$\Re\sigma_V(p)=\Re\frac{A(p)}{p^2A^2(p)+B^2(p)}$ in the complex plane of the external momentum without contour deformation. \subref{fig:2_a}: IR part only. \subref{fig:2_b}: Full Maris-Tandy model \cite{Maris:1999nt}.}
\label{fig:2}
\end{figure*}
Figure \ref{fig:2} shows the solution without \ref{fig:2_a} and with \ref{fig:2_b} UV term. At first sight it appears that the inclusion of the UV term only has a mild effect on the solution. However, one has to keep in mind that the solution shown in Figure \ref{fig:2_b} has been obtained without contour adjustment, and is thus not the actual solution of the quark propagator in the complex plane. Also, upon close inspection (but unfortunately hard to see in print), the surface of this plot is much more rugged than the smooth surface of the IR solution shown in \ref{fig:2_a}. This is a typical sign that the integration contour along the real axis has crossed a branch cut or encountered some  non-analytic obstructions. The poles appear to have shifted slightly. Another interesting aspect is the application of Cauchy's argument principle to the denominator of the vector part of the propagator, as discussed in \cite{Windisch:2016iud}. If only the IR part of the MT model is considered, the integral yields $\delta=N_Z-N_P=3$ as the difference between zeros and poles in the denominator, which gives rise to three poles in the propagator (two c.c. poles, one on the real axis). Once the UV term is added, the same integral changes to two, which can either be due to an additional pole in the expression, or due to the numbers of zeros having decreased. In any case, the resulting analytic properties will be altered, which emphasizes the necessity of a thorough analysis and a proper treatment of the full model.

\section{Acknowledgments}
I thank Pedro Bicudo and all other organizers of \textit{Excited QCD 2017} for this wonderful conference, and also Gernot Eichmann for discussions. Furthermore, I would like to express my gratitude towards my advisor Mark G. Alford, and I also thank David Hall, Sai Iyer and Richard Schmaeng from the Physics Department at Washington University in St.~Louis for providing the computational resources in form of a GPU cluster node that made these calculations possible. Support through the U.S. Department of Energy, Office of Science, Office of Nuclear Physics under Award Number \#DE-FG-02-05ER41375, as well as through the Austrian Science Fund (FWF), Schr\"odinger Fellowship J3800-N27 is acknowledged.

\bibliographystyle{utphys_mod}
\bibliography{literature_eQCD2017}

\end{document}